\documentstyle[pra,aps]{revtex}
\begin{document}
\draft

\title{Non-deterministic Chaos}

\author{D. D. Dixon}
\address{
Institute for Geophysics and Planetary Physics\\
University of California \\
Riverside, CA 92521\\
DIXON@UCRPHF.UCR.EDU
}

\date{\today}

\maketitle

\begin{abstract}
Non-deterministic chaos is a new dynamical paradigm where a non-deterministic
system is influenced by random perturbations to produce the
appearance of complexity.  The non-determinism is envisioned
to occur only at a single point in phase space, where many trajectories
intersect.  In the presence of external random
perturbations (noise), whenever the phase space trajectory approaches the
singularity, it will jump in an unpredictable way to a different
solution.  This behavior, while similar in appearance to deterministic
chaos, has rather different implications for prediction and control.
\end{abstract}

\pacs{05.40.+j, 05.45+b}

\section*{Introduction}

It is often assumed that classical physical systems are
{\em deterministic}~\cite{earman}, meaning that the forward time evolution
is uniquely determined by the current state of the system.
Correspondingly, determinism also implies that the system's current
state uniquely determines its past behavior.  These qualities follow
directly from the {\em Existence and Uniqueness Theorem} of differential
equations~\cite{coddington}, which requires that the equations of
motion of the system are Lipschitz continuous.  However, this is
nothing in classical mechanics that {\em requires} Lipschitz continuity.
Indeed, in the case of a cracking whip, the physical solutions imply
violation of the Lipschitz condition~\cite{zak1}.  A similar effect
is seen in seismic waves as they approach the surface of the
Earth~\cite{zak2}.  If Nature is not required to be Lipschitzian,
we must ask if it required to be deterministic.  Some types of
non-determinism in classical systems have been explored
previously~\cite{zak3,chen}.  In this work, we examine a somewhat
different flavor of non-determinism, one which implies the possible
existence of a new dynamical behavior: {\em non-deterministic chaos}.\\

\section*{The Non-deterministic Harmonic Oscillator (NDHO)}

We begin with the simple harmonic oscillator (SHO), described by the
equations
\begin{eqnarray}\label{eq:sho}
\frac{d}{dt}x & = & y\nonumber\\
\frac{d}{dt}y & = & -x.
\end{eqnarray}
Solutions of eqs.~\ref{eq:sho} are circles in the $(x,y)$ phase plane.
The SHO is a deterministic system, so that every point $(x,y)$ belongs
to a unique solution described by a circle with a particular
radius $r = \sqrt{x^2 + y^2}$.\\

Suppose, now, that we apply the following non-linear
coordinate transformation to the SHO phase space:
\begin{equation}
x \rightarrow x - r = x - \sqrt{x^2 + y^2}.
\end{equation}
This translates all points on a circle of radius $r$ in
the positive $x$-direction by an amount equal to $r$.
A family of circles concentric about the origin in the original
space will now share a common tangent point at the origin of the
transformed space (see Figure~\ref{fig:circles}).\\

The equation of a transformed circle in the new space is given
by
\begin{equation}
(x - r)^2 + y^2 = r^2,
\end{equation}
or, solving for $r$,
\begin{equation}\label{eq:r}
r = \frac{1}{2 x}(x^2 + y^2).
\end{equation}
Using eqn.~\ref{eq:r}, we can easily apply the transformation to
the SHO.  The transformed SHO equations of motion in the new
coordinate system are given by
\begin{eqnarray}\label{eq:ndho}
\frac{d}{dt}x & = & y\nonumber\\
\frac{d}{dt}y & = & \frac{y^2}{2 x} - \frac{1}{2}x.
\end{eqnarray}
{}From the above discussion, the solutions of eqs.~\ref{eq:ndho}
will be the family of transformed circles, all sharing a common
tangent point at the origin.  Such intersection of many phase space
trajectories is not so unusual.  An attracting fixed point, for
example, is approached asymptotically for all initial conditions
in its basin of attraction (i.e., solutions are unique for finite
times).  What {\em is} unusual about eqs.~\ref{eq:ndho} is that
the common point is intersected in finite time, and further
is not a fixed point.  This is easily seen by taking the limit
of eqs.~\ref{eq:ndho} along a solution of radius $r$:
\begin{eqnarray}\label{eq:limit}
\lim_{x,y\rightarrow 0} \frac{d}{dt}x & = & 0,\nonumber\\
\lim_{x,y\rightarrow 0} \frac{d}{dt}y & = & r.
\end{eqnarray}
Thus, the origin is a singularity of eqs.~\ref{eq:ndho}, where neither
past nor future time evolution is uniquely determined.  Henceforth,
we shall refer to eqs.~\ref{eq:ndho} as the {\em non-deterministic
harmonic oscillator} (NDHO).\\

The NDHO provides the paradigm for the type of system we are examining:
solutions of the equations of motion are a family of closed loops
(``transients'') all sharing a common tangent point.  From
eqs.~\ref{eq:limit}, the dynamics of the NDHO are not defined
by the equations of motion alone
(this is not a necessary condition for a system to be
non-deterministic~\cite{dixon2}).
However, let us imagine we built an NDHO in a laboratory.  How would
it behave?  We note that all {\em physical} systems are subject to
external perturbations, or ``noise''.  While the physical state of
our NDHO is far (in phase space) from the point $(0,0)$, external
noise will have little effect, provided the average amplitude of
the fluctuations is small compared to $r$ for that trajectory.
However, as the trajectory approaches the origin, noise plays a
larger role.  Solutions for all $r$ converge together, ultimately
intersecting at $(0,0)$.  Thus, noise will cause the trajectory
to jump between solutions of widely differing $r$ in a random way.\\

What is the effect of this on our laboratory NDHO?  Suppose we begin
the system on a solution of radius $r_1$.  As we watch the system
evolve forward in time, we will find that after it passes near the
origin the trajectory has changed to a completely different solution
of radius $r_2$.  Repeating the experiment with the same initial
conditions, we find that the trajectory jumps to a completely
different solution of radius $r_3$, where $r_3 \neq r_2$.  Were we
to repeat this a large number of times, for different values of
$r_1$, we would find that the
solution after the singularity is completely unrelated to the
solution before.  If the NDHO were allowed to run for several
oscillations, a time series measurement of one variable would
appear as a piecewise continuous sequence of oscillations with
different amplitudes.  Further, the sequence of amplitudes would
be random and unpredictable.  We
term this behavior {\em non-deterministic chaos}, non-deterministic
because its origin lies in the non-determinism at a non-Lipschitz
singularity, and chaos because of the long term unpredictability
of the dynamics.\\

\section*{A Physically Motivated Example}

The NDHO, while illustrative of the type of non-determinism we
are examining, is also a somewhat contrived
example.  We now describe a non-deterministic system
based on physical considerations.  This system is a model of the
behavior of neutron star magnetic fields.  We describe it briefly;
for a more detailed discussion, the reader is referred to~\cite{cummings}.\\

The model envisions two oppositely charged spherical shells which are
allowed to rotate differentially.  The magnetization of one shell
(${\bf M}_1$) will interact with the magnetic field of the second shell
(${\bf H}_2$), as well as experience non-electromagnetic (``mechanical'')
interactions with the surrounding medium.  The magnetic interactions
include a term to induce precession of ${\bf M}_1$ about the instantaneous
direction of ${\bf H}_2$, and the Landau-Lifshitz magnetic damping,
which tends to align ${\bf M}_1$ with the direction of ${\bf H}_2$.  The
mechanical interaction is taken as a simple damping, proportional
to the difference in angular velocities of the two spheres.  Parameterizing
the interactions, we obtain the following equations:
\begin{equation}\label{eq:eqmotion}
\frac{d}{dt}{\bf M}_1 =
\bar{\gamma}({\bf M}_1 \times {\bf H}_2)
- \bar{\lambda}(\frac{{\bf M}_1 \cdot {\bf H}_2}{M_1^2}{\bf M}_1 - {\bf H}_2)
- \overline{\overline{\eta}} \cdot ({\bf \omega}_1 - {\bf \omega}_2).
\end{equation}
Following the scaling procedure described in~\cite{cummings}, and
conserving angular momentum, we arrive at the following equation:
\begin{equation}\label{eq:model}
\frac{d}{d\tau}{\bf m} = - {\bf m} \times \hat{z}
- \lambda\left(\frac{{\bf m} \cdot \hat{z}}{m^2}{\bf m} - \hat{z}\right)
- \overline{\overline{\epsilon}} ({\bf m} - \hat{z}),
\end{equation}
where ${\bf m}$ is the scaled magnetization, $\tau$ is the scaled
time, $\lambda$ is the scaled
Landau damping parameter, and $\overline{\overline{\epsilon}}$ is the
scaled viscous damping parameter tensor.\\

Examination of eqn.~\ref{eq:model} reveals axial symmetry about the
$z$-axis.  This prompts us to make the following transformation:
\begin{eqnarray}
\nonumber x & = & \sqrt{{m_x}^2 + {m_y}^2} \\
z & = & m_z \\
\nonumber \phi & = & \arctan{\frac{m_y}{m_x}}
\end{eqnarray}
which implies
\begin{eqnarray}\label{eq:xform}
\nonumber m_x & \rightarrow & x\cos{\phi} \\
m_y & \rightarrow & x\sin{\phi} \\
\nonumber m_z & \rightarrow & z
\end{eqnarray}
Substituting the above transformations into eqn.~\ref{eq:model}, we
obtain
\begin{eqnarray}\label{eq:2d}
\nonumber \dot{x} & = & \frac{\lambda x z}{x^2 + z^2} - \epsilon x \\
\dot{z} & = & \frac{\lambda z^2}{x^2 + z^2} - \bar{\epsilon} z
                - (\lambda - \bar{\epsilon})
\end{eqnarray}
and
\begin{equation}
\dot{\phi} = 1
\end{equation}
where an overdot again represents differentiation with respect to the
scaled time $\tau$.  The $\phi$ equation is trivial, simply representing
a constant precession about the $z$-axis.  Any interesting dynamical
behavior must occur in eqs.~\ref{eq:2d}.\\

Numerical integration of eqs.~\ref{eq:2d} for $\epsilon < \lambda$
yields the phase space plot in Figure~\ref{fig:nstar}, and the
time series in Figure~\ref{fig:nstartime}.
Note the apparent intersection of trajectories at the origin, indicating
possible non-determinism at that point.  Indeed, it can be
shown~\cite{dixon2,dixon3} that for the RHS of eqs.~\ref{eq:2d},
the point $(0,0)$ is a non-Lipschitz singularity.  Given this, the
non-deterministic nature of eqs.~\ref{eq:2d} at $(0,0)$ is proved
by the following argument (for more details the reader is referred
to~\cite{dixon2,dixon3}):
\begin{enumerate}
\item{The only fixed point of eqs.~\ref{eq:2d} for $\epsilon < \lambda$
is at $(0,1)$.}
\item{The sole fixed point is a saddle, and thus is not an attractor
for a set of initial conditions of non-zero measure~\cite{verhulst}.
This also implies that no periodic orbits exist~\cite{verhulst}.}
\item{By the Poincar\'e-Bendixson Theorem, if the solutions of a
two-degree of freedom dynamical system contain only Lipschitz points,
then the only possible bounded asymptotic solutions are
stationary or periodic.
As neither possibility exists for eqs.~\ref{eq:2d}, $(0,0)$ must
be contained by {\em all} bounded solutions, and therefore this point
is non-deterministic.}
\end{enumerate}

\section*{Non-determinism and predictability}
Non-deterministic chaos has the property of being predictable for
short times (between intersections of the singularity), yet
completely unpredictable over long time periods.  Long-term
unpredicability is also one of the hallmarks of {\em deterministic}
chaos, but it is here that the similarity ends.  Aside from being
described by deterministic equations, deterministic
chaos is often characterized by exponential divergence of initially close
solutions, and associated with an complex fractal structure,
the strange attractor.  Non-deterministic chaos derives its
unpredictability from a more violent, but localized instability.
Further, there exists no attractor, strange or otherwise, at least
in the usual sense of the word.\\

Figures~\ref{fig:movie} illustrate the loss of information at
the singularity due to the presence of external noise.  A set of
initial conditions in evolved via eqs.~\ref{eq:2d}.  Initially,
we see the phase space stretching, similar to what occurs in
deterministic chaos.  However, once the singularity is encountered,
points are scattered, and soon are randomly spread through a
region of phase space (in this case, the region is enclosed by the
homoclines of the saddle at (0,1)~\cite{dixon2}).  This behavior
is in stark contrast to what one expects from deterministic chaos,
where an initial volume is stretched and folded~\cite{rasband},
spreading across the attractor in a smooth fashion.  Dynamical
measures such as the Lyapunov exponent are meaningless.  The
``attractor'' exists only in a statistical sense, representing
the probability density that a particular point in phase space
will be visited.  Formally, one could find this probability density
by transforming eqs.~\ref{eq:2d} into a Brownian motion and
numerically solving the forward Kolmogorov equation~\cite{bhattacharya}.
This approach is complicated by the existence of the singularity.
Instead, we simply integrated eqs.~\ref{eq:2d} for 200 million time
steps, and constructed the distribution from this.  The results are
shown in Figure~\ref{fig:probdist}.\\

Figure~\ref{fig:probdist} gives a long-term, global statistical
picture.  However, the nature of non-deterministic chaos allows
us to easily extract more useful statistical information.  In
particular, we shall utilize the fact that away from the singular
point, the dynamics is quite well-behaved.
Let us return to the NDHO, as its solutions are known analytically.
The solutions of the NDHO may be parameterized by their radius.
Solutions away from the singularity have essentially constant
radii; the big jumps occur only near the singularity.
Can we predict the probability that a circle of given $r$ is chosen when
the orbit leaves some neighborhood about the origin?  Let us define
this neighborhood as a disk of radius $\delta$, and note that an
orbit leaving this neighborhood does so with angle $\theta$, which
we take as measured from the $y$-axis.  Now, assuming the external
fluctuations to be isotropic, the probability density of picking
a particular $\theta$ is constant, i.e.,
\begin{equation}
p(\theta)d\theta \propto d\theta.
\end{equation}
Next, we note that everywhere
except at the origin the Existence and Uniqueness Theorem applies to
solutions of eqs.\ (\ref{eq:ndho}), thus each circle of radius
$r$ is associated with a unique $\theta$, and we may write $\theta$
as a function of $r$.  Substituting into $p(\theta)$, we find
\begin{equation}
p(r)dr \propto \frac{\partial \theta(r)}{\partial r}dr.
\end{equation}
The probability of getting a circle between $r$ and $r + \Delta r$
is simply
\begin{equation}
P(r,\Delta r) \propto
\int_{r}^{r+\Delta r}\frac{\partial \theta(r)}{\partial r}dr
 = \theta(r+\Delta r) - \theta(r).
\end{equation}
For the case at hand, we find
\begin{equation}
P(r,\Delta r) \propto \arccos{\frac{\delta}{2(r+\Delta r)}}
                        - \arccos{\frac{\delta}{2r}}.
\end{equation}

This approach (first described in~\cite{dixon3}) is somewhat
simplified.  A rigorous derivation would account for the statistical
properties of the noise, and derive $P(r, \Delta r)$ via
stochastic calculus.  The above does show, however, that the
simple structure of the solutions of a non-deterministic system
lends itself to the construction of statistical arguments.
Further, with a judicious choice of $\delta$, based on knowledge of the
average amplitude of the fluctuations, the above procedure should
yield a good approximation of the true distribution $P(r,\Delta r)$.\\

\section*{Controlling Non-deterministic Chaos}

The control of deterministic chaotic systems using small perturbations
has been a subject of recent vigorous research~\cite{shinbrot}.
The most popular method of controlling deterministic chaos involves
the stabilization of (otherwise) unstable periodic orbits which
are embedded in the chaotic motion.  As there exist an infinity of
orbits, a rich variety of behaviors may be extracted from the
controlled deterministic chaotic system, allowing for flexibility
and easy optimization of a system's behavior.\\

For a non-deterministic chaotic system, we have a similar situation.
With a continuum of different solutions intersecting at a single
point, we can easily effect control via an appropriate perturbation.
Similar to the previous section, we simply examine how solutions
leave a $\delta-$neighborhood about the singularity.  Again, away from the
singular point the solution is well-defined.  Suppose that
each different solution may be parameterized by some quantity
$\gamma$ (in the case of the NDHO, this is the radius).  A given
solution, parameterized by $\gamma_0$, will intersect the
$\delta$-neighborhood at a unique point $(x_0,y_0)$.  From this,
we may construct the angle $\theta(\gamma_0) =
\arctan{x_0/y_0}$.\\

The angle $\theta(\gamma)$ we term the {\em control angle}, and the
reason should be obvious.  To keep the system on a solution with
parameter $\gamma_0$, we need only to wait until the trajectory
approaches the origin, and then perturb it so that it leaves at
angle $\theta(\gamma_0)$.  This perturbation will be quite small,
of the order of $\delta$.  with the size of $\delta$ being determined
largely by the noise amplitude.  We see that in a non-deterministic
system, there is a continuum of possibilities available through
small control perturbations.  If a change in system behavior were
required, it is easily and quickly effected by simply changing
$\theta(\gamma)$.  In fact, one could vary $\theta(\gamma)$ as
a function of time to induce arbitrarily complex behavior.\\

As an example, we have applied this control algorithm to
eqs.~\ref{eq:2d}.  To simulate the effect of noise, a small
($10^{-4}$ of the integration stepsize) normally distributed random
number was added at each integration step.  The controlled
signals for various values of $\theta$ are shown in
Figure~\ref{fig:control}.  Figure~\ref{fig:contnoise} shows
the effect of noise for different values of $\delta$.\\

\section*{Discussion}

The type of ``non-determinism'' described above should not be
construed as implying stochasticity.  Indeed, the behavior of
both the NDHO and neutron star model are uniquely determined
away from the singular point.  It is at this point, and this
point only, that the non-deterministic nature of the equations
arises.  In the presence of random fluctuations, which are
ubiquitous (though perhaps small) in physical systems, the
non-determinism, albeit it at a single point, becomes important.
The resulting dynamics, which we have termed {\em non-deterministic
chaos}, consist of a random sequence of ``transient'' oscillations.\\

This work does not represent the first suggestion that non-determinism
exists in classical systems.  The {\em terminal dynamics} described
by M. Zak~\cite{zak3} utilizes a similar mechanism where multiple
trajectories intersect at a common equilibrium point in finite time.
Chen~\cite{chen} has independently suggested the same behavior
under the heading of {\em noise induced instability} (though
non-determinism as such is never explicitly mentioned).  For
clarity, we shall henceforth refer to terminal dynamics as
``Type I non-determinism'', and the dynamics described herein
as ``Type II non-determinism''.  The primary difference between
the two is that for Type I non-determinism, the singularity
occurs at an equilibrium point of the equations of motion, while
for Type II, the singularity is shared among a group of
dynamic trajectories.  The physical implications of this
difference are yet to be explored.  Other aspects of non-deterministic
systems, especially in the presence of noise, have been explored
by A. H\"ubler~\cite{hubler}.\\

There is no principle in nature that precludes the existence of
the types of systems we have described.  Indeed, an analog
VLSI circuit has been built which displays Type I
non-determinism\cite{cetin}.  The neutron star model displays
Type II non-determinism, and is based on perfectly reasonable
physical assumptions.  Work is in progress to build an electronic
analog of this model.
It has also been suggested that non-determinism may play an
important role in biological systems~\cite{zbilut}.
However, the question arises as to the ubiquity of such systems.
Is non-determinism a generic property of
Nature, or have we simply stumbled upon a few pathological
examples?  We cannot at this time answer that question.\\

However, we feel further investigation is warranted.  In particular,
application of several standard measures (power spectrum,
Lyapunov exponent, etc.) to a time series generated by eqs.~\ref{eq:2d}
would lead one to believe that one examining an instance
of {\em deterministic} chaos~\cite{dixon2}.  As we have seen, though,
issues of prediction and control would be addressed much differently
for a non-deterministic chaotic system.  Indeed, Crutchfield has
shown that in the context of model building, assuming determinism
when the underlying process is non-deterministic leads
to undue complexity in the model~\cite{crutchfield}.  It would
seem reasonable to search for non-deterministic chaos in
apparently complex systems, especially in cases where traditional
analysis tools (which, again, assume determinism) have failed.
In a forthcoming paper, we will address the problem of detecting
non-determinism in observed data.\\

\begin{figure}
\caption{Some examples of circular orbits of different radii, all sharing
a common point at the origin.}
\label{fig:circles}
\end{figure}

\begin{figure}
\caption{Phase plot of solutions in the neutron star model for
$\lambda=1.0,\epsilon=0.6,\overline{\epsilon}=0.7$.}
\label{fig:nstar}
\end{figure}

\begin{figure}
\caption{Time series $z(t)$ vs. $t$ for the neutron star model with
parameter values $\lambda=1.0,\epsilon=0.6,\overline{\epsilon}=0.7$.}
\label{fig:nstartime}
\end{figure}

\begin{figure}
\caption{Loss of information due to the singularity. a) 10,000 initial
points are arranged in a 100$\times$100 square. b) Initial evolution.
c) When the singularity is encountered, initally close points are
scattered randomly. d) All information about the initial conditions
has been lost.  The only information carried by the system is in the
density of trajectories.}
\label{fig:movie}
\end{figure}

\begin{figure}
\caption{Probability of finding the system in a particular
region of phase space.  The distribution was found by integrating
the equations for 200 million steps and totalling the amount of time
spent in each region.}
\label{fig:probdist}
\end{figure}

\begin{figure}
\caption{Examples of output from the neutron star model when the
control algorithm is applied.  Signals are shown for
$\theta(\gamma) = $ a) 0.005, b) 0.03, c) 0.08, and d) 0.2.}
\label{fig:control}
\end{figure}

\begin{figure}
\caption{The control algorithm begins to break down if $\delta$ is
chosen to be comparable to the noise level.  Signals are shown
for $\delta =$ a) $10^4 \sigma$, b) $10^3 \sigma$, c) $10^2 \sigma$,
and d) $10 \sigma$, where $\sigma$ is the RMS of the noise.}
\label{fig:contnoise}
\end{figure}


\begin{references}
\bibitem{earman}J. Earman, {\it A Primer on Determinism},
Reidel Publishing Corp., Dordrecht, Holland, 1986.
\bibitem{coddington}E. A. Coddington and N. Levinson,
{\it Theory of Ordinary Differential Equations}, McGraw-Hill,
New York, 1955.
\bibitem{zak1}M. Zak, PMM {\bf 39}, 1048 (1970).
\bibitem{zak2}M. Zak, J. Applied Mech. {\bf 50}, 227 (1983).
\bibitem{zak3}M. Zak,
International J. of Theoretical Physics, {\bf 32}, 159, (1993).
\bibitem{chen}Z.-Y. Chen, Phys.\ Rev.\ A {\bf 42}, 5837 (1990).
\bibitem{dixon2}D. D. Dixon, F. W. Cummings, and P. E. Kaus,
Physica D {\bf 65}, 109 (1993).
\bibitem{cummings}F. W. Cummings, D. D. Dixon, and P. E. Kaus,
Astrophys.\ J. {\bf 386}, 215 (1992).
\bibitem{dixon3}D. D. Dixon, {\it A Theoretical Model of Gamma-Ray
Bursts}, Ph.D. Thesis, University of California, Riverside (1993).
\bibitem{verhulst}F. Verhulst, {\it Nonlinear Differential Equations and
Dynamical Systems}, Springer-Verlag, Berlin, 1990.
\bibitem{rasband}S. N. Rasband,{\it Chaotic Dynamics of Nonlinear
Systems}, John Wiley \& Sons, New York, 1990.
\bibitem{bhattacharya}R. N. Bhattacharya and E. A. Waymire,
{\it Stochastic Processes with Applications}, Wiley, New York, 1990.
\bibitem{shinbrot}T. Shinbrot, C. Grebogi, E. Ott, and J. A. Yorke,
Nature {\bf 363}, 411 (1993).
\bibitem{hubler} A. H\"ubler, in {\em Modeling Complex Phenomena},
edited by L. Lam and V. Naroditsky, Springer, New York, 1992.
\bibitem{cetin}B. C. Cetin, D. A. Kerns, J. W. Burdick, and J. Barhen,
Robotics and Mechanical Systems Report No. RMS-92-01, Department
of Mechanical Engineering, California Institute of Technology, Pasadena,
1991.
\bibitem{zbilut}J. P. Zbilut, A. H\"ubler, C. A. Webber Jr.,
in {\it Fluctuations and Order: The New Synthesis},
edited by M. Millonas, Springer, in press.
\bibitem{crutchfield}J. P. Crutchfield, in {\it Inside versus Outside},
edited by H. Atmanspacher, Series in Synergetics,
Springer-Verlag, Berlin, 1993.
\end{references}
\end{document}